\documentstyle[epsfig]{mn}

\title[Random-Walk Statistics]{Random-Walk Statistics and the Spherical Harmonic Representation of CMB Maps}

\author[Andrew Stannard \&  Peter Coles]{Andrew Stannard \& Peter Coles\\
School of Physics \& Astronomy, University of Nottingham,
University Park, Nottingham, NG7 2RD, United Kingdom\\ }

\begin{document}

\maketitle

\begin{abstract}
We investigate the properties of the (complex) coefficients
obtained in a spherical harmonic representation of temperature
maps of the cosmic microwave background (CMB). We study the effect
of the coefficient phase only, as well as the combined effects of
phase and amplitude. The method used to check for anomalies is to
construct a ``random walk'' trajectory in the complex plane where
the step length and direction are given by the amplitude and phase
(respectively) of the harmonic coefficient. If the fluctuations
comprise a homogeneous and isotropic Gaussian random field on the
sky, the path so obtained should be a classical ``Rayleigh
flight'' with very well known statistical properties. We
illustrate the use of this random-walk representation by using the
net walk length as a test statistic, and apply the method to the
coefficients obtained from a Wilkinson Microwave Anisotropy Probe
(WMAP) preliminary sky temperature map.
\end{abstract}


\section{Introduction}
The process of gravitational instability initiated by small
primordial density perturbations is a vital ingredient of
cosmological models that attempt to explain how galaxies and
large-scale structure formed in the Universe. In the standard
cosmological models, a period of accelerated expansion
(``inflation'') generated density fluctuations with simple
statistical properties through quantum processes (Starobinsky
1979, 1980, 1982; Guth 1981; Guth \& Pi 1982; Albrecht \&
Steinhardt 1982; Linde 1982). In this scenario the primordial
density field is assumed to form a statistically homogenous and
isotropic Gaussian Random Field (GRF). Over years of observational
scrutiny this paradigm has strengthened its hold in the minds of
cosmologists and has survived many tests, culminating in those
furnished by the Wilkinson Microwave Anisotropy Probe (WMAP;
Bennett et al. 2003; Hinshaw et al 2003).

Since the release of the first (preliminary) WMAP data set it has
been subjected to a number of detailed independent analyses that
have revealed some surprising features. Eriksen et al. (2004) have
pointed out the existence of a North-South asymmetry suggesting
that the cosmic microwave background (CMB) revealed by the WMAP
data is not statistically homogeneous over the celestial sphere.
This is consistent with the results of Coles et al. (2004) who
found evidence for phase correlations in the WMAP data; see also
Hajian \& Souradeep (2003) and Hajian, Souradeep \& Cornish
(2004). The low--order multipoles of the CMB also display some
peculiarities (de Oliveira-Costa et al. 2004a; Efstathiou 2004).
Vielva et al. (2004) found significant non--Gaussian behaviour in
a wavelet analysis of the same data, as did Chiang et al. (2004),
Larson \& Wandelt (2004) and Park (2004). Other analyses of the
statistical properties of the WMAP have yielded results consistent
with the standard form of fluctuation statistics (Komatsu et al.
2003; Colley \& Gott 2003). These unusual properties may well be
generated by residual foreground contamination (Banday et al.
2003; Naselsky et al. 2003; de Oliveira-Costa et al. 2004; Dineen
\& Coles 2004) or other systematic effects, but may also provide
the first hints of physics beyond the standard cosmological model.

In order to tap the rich source of information provided by future
CMB maps it is important to devise as many independent statistical
methods as possible to detect, isolate and diagnose the various
possible causes of departures from standard statistics. One
particularly fruitful approach is to look at the behaviour of the
complex coefficients that arise in a spherical harmonic analysis
of CMB maps. Chiang et al. (2004), Chiang, Naselsky \& Coles
(2004), and Coles et al. (2004) have focussed on the phases of
these coefficients on the grounds that a property of a
statistically homogenous and isotropic GRF is that these phases
are random. Phases can also be use to test for the presence of
primordial magnetic fields (Chen et al. 2004; Naselsky et al.
2004) or evidence of non-trivial topology (Dineen, Rocha \& Coles
2004). In this paper we suggest an extension of this idea which
involves viewing the entire set of harmonics (both amplitude and
phase) in the complex plane.

We describe the basic harmonic description in the next Section. In
Section 3 we show how to represent the spherical harmonics as
random-walks in the complex plane and give some simple analytic
results. We apply the idea to the preliminary WMAP 1-year data in
Section 4 and briefly discuss the results in Section 5.

\section{Spherical Harmonics and Gaussian Fluctuations}

We can describe the distribution of fluctuations in the microwave
background over the celestial sphere using a sum over a set of
spherical harmonics:
\begin{equation}
\label{deltatovert} \Delta (\theta, \phi)= \frac{T(\theta ,\phi
)-\bar{T}}{\bar{T}}=\sum _{l=1}^{\infty }\sum _
{m=-l}^{m=+l}a_{l,m}Y_{lm}(\theta ,\phi ).
\end{equation}
Here $\Delta(\theta,\phi)$ is the departure of the temperature
from the average at angular position $(\theta ,\phi)$ on the
celestial sphere in some coordinate system, usually galactic. The
$Y_{lm}(\theta ,\phi)$ are spherical harmonic functions which we
define in terms of the Legendre polynomials $P_{lm}$ using
\begin{equation}
Y_{lm}(\theta ,\phi )= (-1)^m
\sqrt{\frac{(2l+1)(l-m)!}{4\pi(l+m)!}} P_{lm}(\cos \theta) {\rm e}
^{im\phi},
\end{equation}
i.e. we use the Condon-Shortley phase convention. In Equation (1),
the $a_{l,m}$ are complex coefficients which can be written
\begin{equation}
a_{l,m}=x_{l,m} + i y_{l,m} = |a_{l,m}|\exp[i\phi_{l,m}].
\end{equation}
Note that, since $\Delta$ is real, the definitions (2) \& (3)
requires the following relations between the real and imaginary
parts of the $a_{l,m}$: if $m$ is odd then
\begin{eqnarray}
x_{l,m}= \Re (a_{l,m}) &  = & - \Re(a_{l,-m})=-x_{l,-m},
\nonumber\\ y_{l,m}= \Im(a_{l,m}) &  = & \Im(a_{l,-m}) = y_{l,-m};
\end{eqnarray}
while if $m$ is even
\begin{eqnarray}
x_{l,m}=\Re (a_{l,m}) & = & \Re(a_{l,-m})=x_{l,-m}, \nonumber\\
y_{l,m}=\Im(a_{l,m}) & = & - \Im(a_{l,-m})=y_{l,-m};
\end{eqnarray}
and if $m$ is zero then
\begin{equation}
\Im(a_{l,m}) =y_{l,0} = 0.
\end{equation}
From this it is clear that the $m=0$ mode always has zero phase,
and there are consequently only $l$ independent phase angles
describing the harmonic modes at a given $l$. Without loss of
information we can therefore restrict our analysis to $m\geq 0$.

If the primordial density fluctuations form a Gaussian random
field in space the temperature variations induced across the sky
form a Gaussian random field over the celestial sphere. This means
that
\begin{equation}
\langle a_{l,m}a_{l',m'}^* \rangle = C_l \delta_{ll'}\delta_{mm'},
\end{equation}
where $C_l$ is the angular power spectrum, the subject of much
scrutiny in the context of the cosmic microwave background (e.g.
Hinshaw et al. 2003), and $\delta_{xx'}$ is the Kronecker delta
function. Since the phases are random, the stochastic properties
of a statistically homogeneous and isotropic Gaussian random field
are fully specified by the $C_l$, which determines the variance of
the real and imaginary parts of $a_{l,m}$ both of which are
Gaussian:
\begin{equation}
\sigma^{2} (x_{l,m}) = \sigma^{2}(y_{l,m})= \sigma_l^2=
\frac{1}{2} C_l.
\end{equation}

\section{Random Walks in Harmonic Space}

To begin with, we concentrate on a simple measure based on the
distribution of total displacements. Consider a particular value
of $l$. The set of values $\{{\bf a_{l,m}}\}$ can be thought of as
steps in a random walk in the complex plane, a structure which can
be easily visualized and which has well-known statistical
properties.

The simplest statistic one can think of to describe the set
$\{{\bf a_{l,m}}\}$ is the net displacement of a random walk
corresponding to the spherical harmonic mode $l$, i.e.
\begin{equation}
{\bf R_l} \equiv (X_l, Y_l) = \sum_{m>0} {\bf a_{l,m}},
\end{equation}
where the vector ${\bf a_{l,m}} \equiv (x_{l,m},y_{l,m})$ and the
random walk has an origin at $a_{l,0}$ (which is always on the
$x$-axis). The length of each step $a_{l,m}=|{\bf a_{l,m}}|$ is
the usual spherical harmonic coefficient described in the previous
section and defined by equation (1). If the initial fluctuations
are Gaussian then the two components of each displacement are
independently normal with zero mean and the same variance (8).
Each step then has a Rayleigh distribution so that the probability
density for $a_{l,m}$ to be in the range $(a,a+da)$ is
\begin{equation}
p(a)=\frac{a}{\sigma_l^2}
\exp\left(-\frac{a^2}{2\sigma_l^2}\right).
\end{equation}
This is a particularly simple example of a random walk (McCrea \&
Whipple 1940; Chandrasekhar 1943;  Hughes 1995). Since the
displacements in $x$ and $y$ are independently Gaussian the next
displacement after $l$ steps is itself Gaussian with variance
$l\sigma_l^2$. The total displacement $R_l$ is given by
\begin{equation} R_l^2 = \left( X_l^2 + Y_l^2\right).
\end{equation} The probability density of $|R_l|$ to be in the
range $(r,r+dr)$ is then itself a Rayleigh distribution of the
form
\begin{equation}
p_l(r)=\frac{r}{l\sigma_l^2}
\exp\left(-\frac{r^2}{2l\sigma_l^2}\right).
\end{equation}

The result (12) only obtains if the steps of the random walk are
independent and Gaussian. If the distribution of the individual
steps is non-Gaussian, but the steps are independent, then the
result (11) will be true for large $l$ by virtue of the Central
Limit Theorem. Exact results for finite $l$ for example
non-Gaussian distributions are given by Hughes (1995).

A slightly different approach is to keep each step length
constant. The simplest way of doing this is to define
\begin{equation}
{\bf \hat{R}_l} = \sum_{m>0} \frac{{\bf a_{l,m}}}{|{\bf a_{l,m}}|}
= \sum_{m>0} {\bf \hat{a}_{l,m}} ,
\end{equation}
so that each step is of unit length but in a random direction.
This is precisely the problem posed in a famous letter by Pearson
(1905) and answered one week later by Rayleigh (1905). In the
limit of large numbers of steps the result maps into the previous
result (12) with $\sigma_l^2=1$ by virtue of the Central Limit
Theorem. For finite values of $l$ there is also an exact result
which can be derived in integral form using a method based on
characteristic functions (Hughes 1995). The result is that the
probability density for ${\bf \hat{R_l}}$ to be in the range
$r,r+dr$ is
\begin{equation}
q_l(r)= r \int_0^{\infty} u J_0(ur)[J_0(u)]^l du.
\end{equation}
The integral is only convergent for $l>2$ but for $l=1$ or $l=2$
straightforward alternative expressions are available (Hughes
1995).

One can use this distribution to test for randomness of the phase
angles without regard to the amplitudes. To see how this works,
consider the following simple model of phase correlations.
Following Watts, Coles \& Melott (2003), Suppose that the phase
difference between adjacent modes has a preferred angle, as
modelled by the relation
\begin{equation}
{\bf \hat{a}_{l,m}} = {\bf \hat{a}_{l,m+1}} \cos \theta - {\bf
u_{l,m+1}}.
\end{equation}
Here $u_{l,m}$ is a random variable with $|u_{l,m}|=|\sin
\theta|$, $\langle {\bf u_{l,m}} \rangle=0$ and $ {\bf u_{l,m}}
\cdot {\bf \hat{a}_{l,m}}  =0$, to keep the step length equal to
unity. In this model
\begin{equation}
\langle {\bf \hat{a}_{l,m}} \cdot {\bf \hat{a}_{l,m+n}}\rangle =
\langle {\bf \hat{a}_{l,m+1}} \cdot {\bf \hat{a}_{l,m}} \rangle
\cos \theta = \cos^n \theta.
\end{equation}
Hence
\begin{equation}
\langle R_l^2 \rangle = l + 2 \sum_{n>m} \langle {\bf
\hat{a}_{l,m}} \cdot {\bf \hat{a}_{l,m+n}} \rangle,
\end{equation}
which is \begin{equation} \langle R_l^2 \rangle = N + 2
\sum_{n=2}^{l}\sum_{m=1}^{n-1} \cos ^{n-m} \theta.\end{equation}
Using the properties of a geometric series this can be seen to be
\begin{equation}
\langle R_l^2 \rangle = l \left(
\frac{1+\cos\theta}{1-\cos\theta}\right)=\frac{l}{\xi^2},
 \end{equation}
 compared with the simple $\langle R_l^2 \rangle =l$ which obtains
 if the phases are random. The distribution of total displacements
 therefore encodes information about phase correlations in the
 form of a ``persistence length'', $\xi$.
 The distribution of $R_l$ therefore encodes information about
 phase correlations: if there is no preferred direction, $\cos
 \theta=0$ and the result reduces to the case of a purely random
 walk with $\xi=1$. Non-zero values of $\cos \theta$ will alter the mean square
 displacement relative to this case: it can be either larger if
 $\cos\theta$ is positive and steps tend to occur in the same
 direction in the complex plane ($\xi>1$), or smaller if $\xi<1$.

So far we have concentrated on fixed $l$ with a random walk as a
function of $m$. We could instead have fixed $m$ and considered a
random walk as a function of $l$. Or indeed randomly selected $N$
values of $l$ and $m$. In either case the results above still
stand except with $\sigma_l^2$ replaced by an average over all the
modes considered:
\begin{equation}
\sigma^2=\frac{1}{N} \sum_{l,m} \sigma_{l,m}^2.
\end{equation}
We do not consider this case any further in this paper.

\section{Application to the WMAP 1 yr Data}

 There are many possible ways of using the properties of a random walk
 to test the hypothesis that the CMB temperature fluctuations are drawn
from a statistically homogeneous and isotropic Gaussian random
field on the sky could  be furnished by comparing the empirical
distribution of harmonic random flights with the form (12). This
requires an estimate of $\sigma_l^2$. This can either be made
using the same data or by assuming a given form for $C_l$, in
which case the resulting test would be of a composite hypothesis
that the fluctuations constitute a Gaussian random field with a
particular spectrum. For large $l$ this is can be done
straightforwardly, but for smaller values the sampling
distribution of $R_l$ will differ significantly from (12) because
of the uncertainty in population variance from a small sample of
$a_{lm}$. This is the so-called ``cosmic variance'' problem.

\begin{figure}
\centering
 \epsfig{figure=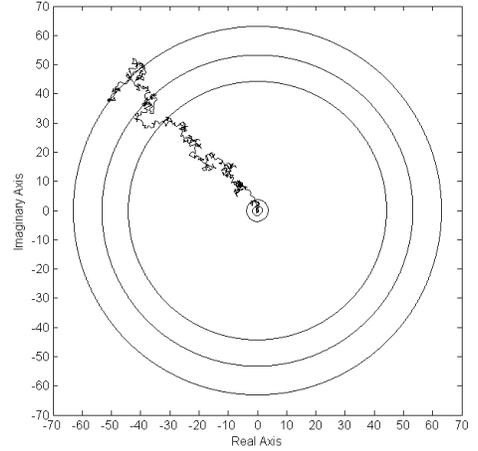,width=11cm,angle=0,clip=}
\caption{\label{fig:fig1} The random walk performed by the
spherical harmonic coefficients for $l=532$, statistically the
mode that displays the greatest departure from that expected under
the null hypothesis. The outer circles correspond to 99.9, 99 and
95 per cent upper confidence limits s (from outer to inner); the
inner circles are the corresponding lower limits, though the 99.9
per cent lower limit is too small to see.}
\end{figure}

In practice the most convenient way to assess the significance of
departures from the relevant distribution would be to perform
Monte Carlo experiments of the null hypothesis. For statistical
measures more complicated than the net displacement, the best way
to set up a statistical test is to use Monte-Carlo re-orderings of
the individual steps to establish the confidence level of any
departure from Gaussianity. This also enables one to incorporate
such complications as galactic cuts.

The WMAP team  released an Internal Linear Combination (ILC) map
that combined  five original frequency band maps in such a way to
maintain unit response to the CMB whilst minimising foreground
contamination. The construction of this map is described in detail
in Bennett et al. (2003). The weighted map is produced by
minimizing the variance of the temperature scale such that the
weights add to one. To further improve the result, the inner
Galactic plane is divided into 11 separate regions and weights
determined separately. This takes account of the spatial
variations in the foreground properties. Thus, the final combined
map does not rely on models of foreground emission and therefore
any systematic or calibration errors of other experiments do not
enter the problem. The final map covers the full-sky and the idea
is that it should represent only the CMB signal. Following the
release of the WMAP 1 yr data Tegmark, Oliveira-Costa \& Hamilton
(2003; TOH) produced a cleaned CMB map. They argued that their
version contained less contamination outside the Galactic plane
compared with the ILC map produced by the WMAP team.

The ILC map is not intended for statistical analysis but in any
case represents a useful ``straw man'' for testing statistical
techniques for robustness. To this end, we analyzed the behaviour
of the random-walks representing spherical harmonic from $l=1$ to
$l=600$ in the WMAP ILC. Similar results are obtained for the TOH
map so we do not discuss the TOH map here. For both
variable-length (9) and unit-length (13) versions of the
random-walk we generated 100000 Monte Carlo skies assuming
Gaussian statistics. These were used to form a distribution of
$|{\bf R_l}|$ (or $|{\bf \hat{R}_l}|$) over the ensemble of
randomly-generated skies. A rejection of the null hypothesis (of
stationary Gaussianity) at the $\alpha$ per cent level occurs when
the measured value of the test statistic lies outstide the range
occupied by $\alpha$ per cent of the random skies. The results we
obtained are summarized in Table 1.

\begin{table}
\begin{center}
\begin{tabular}{|c|c|c|c|}
\hline \hline
 confidence level & 95 & 99 & 99.9 \\
\hline unit length & 33 & 6 & 2 \\ variable length & 20 & 4 & 1
\\
null & 30 & 6 & 0.6\\
 \hline\hline
\end{tabular}
\end{center}
\caption{ \label{tab:tab1}Results for random-walks with variable
step-length and unit step-length for the WMAP ILC data. The
columns show the number of $l$-modes for which the test statistic
lies outside the $\alpha$ percent confidence limit for the Monte
Carlo skies. The final row shows the number of modes that would be
expected, on average, to lie outside this range if the null
hypothesis were true. }
\end{table}

The results show that this simple test does not strongly falsify
the null hypothesis, which is not surprising given the simplicity
of the measure we have used. The number of modes outside the
accepted range is close to that which would be expected if the
null hypothesis were true. Notice that slightly more modes show up
in the unit length case than in the other, perhaps indicating that
the phase correlations that are known to exist in this data
(Chiang et al. 2004) are masked if amplitude information is also
included. The most discrepant mode turns out to be $l=532$ in both
cases. For interest a plot of the random walk for this case is
included as Figure 1.

We can turn these results into limits on the persistence length:
if $\xi$. If $\xi \neq 0$ there would be either more observed
walks longer than the limits ($\xi>1$) corresponding to the null
hypothesis or more shorter ones ($\xi<1$). Assuming the form (12),
which should hold for large values of $l$, the current data place
rough limits of $0.9 < \xi < 1.1$.

\section{Discussion and Conclusions}

We have presented a novel representation of the behaviour of
spherical harmonic coefficients obtained in the representation of
of CMB temperature maps. Indeed, the main result of this paper is
the construction of two-dimensional walks as discussed in Section
3. In order to illustrate the method in the simplest way we
performed a simple statistical test on the net displacement of the
random walks for different $l$ modes. The results of this analysis
do not provide conclusive evidence for departures from a
homogeneous and isotropic GRF. For example, two modes are
discrepant at 99.9 per cent confidence, compared with the 0.6
expected by chance. In a series of independent random skies with
an expectation of 0.6 modes, Poisson statistics give a probability
of 88 per cent that less than two modes would be discrepant. The
detection of two modes is therefore not an indication of a strong
departure from chance.

It is worth noting that the method does not pick out any
significant departures at low $l$, which is where the
non-parametric tests of Coles et al. (2004) were applied with
positive results. This test works better with large numbers of
independent $m$ modes. In the simple form presented here it seems
less powerful than the phase-mapping technique of Chiang et al.
(2003) and Chiang et al. (2004). Whatever forms of non-Gaussianity
and/or non-stationarity there are in these data, this very simple
test is not sensitive to them.

As we explained above, however, the net displacement of the random
walk is a simple but rather crude indication of the properties of
the $\{{\bf a_{l,m}}\}$, as it does not take into account the
ordering of the individual steps.  The possible non-Gaussian
behaviour of the set $\{{\bf a_{l,m}}\}$ is encoded not so much in
the net displacement but in the {\em shape} of the random walk. To
put this another way, there are many possible paths with the same
net displacement, and these will have different shapes depending
on the correlations between step size and direction. Long runs of
directed steps or regular features in the observed structure could
be manifestations of phase correlation (Coles et al. 2004).  For
example, the presence of a primordial magnetic field contribution
to the CMB fluctuations would result in a correlation between
${\bf a_{l-1,m}}$ and ${\bf a_{l+1,m}}$ at fixed $m$ (Chen et al.
2004). Examples of how similar effects might be produced by
foregrounds, galactic cuts and so on are discussed by Naselsky et
al. (2005). These phase correlations would manifest themselves in
higher-order statistics of the random walk as a function of $l$.
It will also be useful, say, to combine random walks as functions
of $m$ for different $l$: correlations between the net
displacements of such walks would indicate the presence of phase
correlations too; see Coles et al. (2004). The graphical
representation of the set $\{{\bf a_{l,m}}\}$ in the form
illustrated by Figure 1 provides an elegant way of visualizing the
behaviour of the harmonic modes and identifying any oddities.
These could be quantified using a variety of statistical shape
measures: moment of inertia (Rudnick, Beldjenna \& Gaspari 1987),
fractal dimension, first-passage statistics, shape statistics
(e.g. Kuhn \& Uson 1982), or any of the methods use to quantify
the shape of minimal spanning trees (Barrow, Bhavsar \& Sonoda
1985). It would be interesting to try these more sophisticated
measures on subsequent releases of the WMAP data.

\section*{Acknowledgments}
We acknowledge the use of the Legacy Archive for Microwave
Background Data Analysis (LAMBDA). Support for LAMBDA is provided
by the NASA Office of Space Science. We are grateful to Patrick
Dineen for supplying us with the spherical harmonic coefficients
for the WMAP ILC. Andrew Stannard received a Mary Cannell summer
studentship while this work was being done.


\begin{thebibliography}{}
\bibitem[Albrecht \& Steinhardt 1982]{as82} Albrecht A., Steinhardt P. J., 1982, Phys. Rev. Lett., 48, 1220
\bibitem[Banday et al. 2003]{bddd03} Banday A.J., Dickinson C., Davies R.D., Davies R.J., G\'orski K.M. 2003,
MNRAS, 345, 897
\bibitem[Bardeen et al. 1986]{bbks} Bardeen J.M., Bond J.R.,
Kaiser N, Szalay A.S., 1986, ApJ, 304, 15
\bibitem[Barrow et al. 1985]{bbs} Barrow J.D., Bhavsar S.P.,
Sonoda D.H., 1985, MNRAS, 216, 17
\bibitem[Bennett et al. 2003]{bhhn03} Bennett C., et al.
2003, ApJS, 148, 1
\bibitem[Chandrasekhar 1943]{c43} Chandrasekhar S., 1943, Rev.
Mod. Phys., 15, 1
\bibitem[Chen et al. 2004]{Chen} Chen G., Mukherjee P.,
Kahniashvili T., Ratra B., Wang Y., 2004, ApJ, 611, 655
\bibitem[Chiang, Naselsky \& Coles 2004]{cnc} Chiang L.-Y.,
Naselsky P.D., Coles P., 2004, ApJ, 602, L1
\bibitem[Chiang et al. 2003]{cnvw} Chiang L.-Y., Naselsky P. D., Verkhodanov O. V., Way M. J.,
2003, ApJ, 590, L65
\bibitem[Coles et al. 2004]{c04} Coles P., Dineen P., Earl J., Wright
D., 2004, MNRAS, 350, 989
\bibitem[Colley \& Gott]{cg03} Colley W.N., Gott J.R., 2003,
MNRAS, 344, 686
\bibitem[De Oliveira-Costa et al. 2004a]{lowl} de Oliveira-Costa
A., Tegmark M., Zaldarriaga M., Hamilton A.J.S., 2004, Phys. Rev.
D., 69, 063516
\bibitem[De Oliveira-Costa et al. 2004b]{forex} de Oliveira-Costa
A., Tegmark M., Davies R.D., Gutierrez C.M., Lasenby A.N., Rebolo
R., Watson R.A., 2004, ApJ, 606, L89
\bibitem[Dineen \& Coles 2004]{dc04} Dineen P., Coles P., 2004, MNRAS, 348, 52
\bibitem[Dineen et al. 2004]{DRC} Dineen P., Rocha G., Coles, P., 2005,
MNRAS, 358, 1285
\bibitem[Efstathiou 2004]{e4} Efstathiou G., 2004, MNRAS, 348, 885
\bibitem[Eriksen et al. 2004]{ehbgl} Eriksen H. K., Hansen F. K., Banday A. J., Gorski K. M., Lilje
P. B., 2004, ApJ, 605, 14
\bibitem[Guth 1981]{g81} Guth A. H., 1981, Phys. Rev. D., 23, 347
\bibitem[Guth \& Pi 1982]{gp82} Guth A. H., Pi S.-Y., 1982, Phys. Rev. Lett. 49, 1110
\bibitem[Hajian \& Souradeep 2003]{hs2003} Hajian A., Souradeep T., 2003, ApJ, 597, L5
\bibitem[Hajian, Souradeep \& Cornish 2004]{hdc} Hajian, Souradeep
T., Cornish N., 2004, preprint, astro-ph/0406354
\bibitem[Hinshaw et al. 2003]{hsvh03} Hinshaw G. et al. 2003, ApJ,
148,
\bibitem[Hughes 1995]{hughes} Hughes B.D., 2004, Random Walks and
Random Environments. Volume 1: Random Walks. Oxford University
Press, Oxford.
\bibitem[Komatsu et al. 2003]{k03} Komatsu E. et al. 2003, ApJ,
148, 119
\bibitem[Kuhn \& Uson 1982]{ku82} Kuhn J.R., Uson J.M., 1982, ApJ,
263, L47
\bibitem[Larson \& Wandelt 2004]{lw04} Larson D.L., Wandelt B.D.,
2004, ApJ, 613, L85
\bibitem[Linde 1982]{l82} Linde, A.D., 1982, Phys. Lett. B., 108, 389
\bibitem[McCrea \& Whipple 1940]{mw40} McCrea W.H., Whipple
F.J.W., 1940, Proc. R. Soc. Edin., 60, 281
\bibitem[Naselsky et al. 2004]{ncov} Naselsky P.D., Chiang L.-Y.,
Olesen P., Verkhodanov O., 2004, astro-ph/0405181
\bibitem[Naselsky, Doroshkevich \& Verkhodanov 2003]{ndv} Naselsky P.D., Doroshkevich A.G., Verkhodanov O., 2003, ApJ,
599, L53
\bibitem[Naseksly et al. 2005]{ncon} Naselsky P.D., Chiang L.-Y.,
Olesen P., Novikov I., 2005, Phys. Rev. D., submitted,
astro-ph/0505011
\bibitem[Park 2004]{p04} Park, C.-G., 2004, MNRAS, 349, 313
\bibitem[Pearson 1905]{pears} Pearson K., 1905, Nature, 72, 294
\bibitem[Rayleigh 1905]{ray} Rayleigh, 1905, Nature, 72, 318
\bibitem[Rudnick et al.]{rud} Rudnick J., Beldjenna A., Gaspari
G., 1987, J. Math. Phys. A, 20, 971
\bibitem[Smoot et al. 1992]{smo92} Smoot G.F. et al. 1992, ApJ, 396, L1
\bibitem[Starobinsky 1979]{s79} Starobinsky A. A., 1979, Pis'ma Zh.
Eksp. Teor. Fiz., 30, 719
\bibitem[Starobinsky 1980]{s80} Starobinsky A. A., 1980 Phys.
Lett. B., 91, 99
\bibitem[Starobinsky 1982]{s82} Starobinsky A. A., 1982, Phys. Lett. B., 117, 175
\bibitem[Tegmark, de Oliveira-Costa \& Hamilton 2003]{TOH03}
Tegmark M., de Oliveira-Costa A., Hamilton A.J.S., 2003, Phys.
Rev. D., 68, 123523
\bibitem[Vielva et al. 2004]{v04} Vielva P., Martinez-Gonzalez E.,
Barreiro R.B., Sanz J.L., Cayon L., 2004, ApJ, 609, 22
\bibitem[Watts et al. 2003]{wcm} Watts P.I.R., Coles P., Melott A.L., 2003, ApJ, 589, L61
\end{thebibliography}
\end{document}